\documentclass[aps, prl, twocolumn, superscriptaddress, floatfix]{revtex4-2}

\usepackage{graphicx}
\usepackage{amsmath}
\usepackage{amssymb}
\usepackage{braket}
\usepackage{hyperref}
\usepackage{xcolor}
\usepackage{soul}
\usepackage{tikz}
\usepackage{float}
\allowdisplaybreaks
\definecolor{linkcolor}{RGB}{0, 0, 255}
\definecolor{citecolor}{RGB}{0, 128, 0}
\definecolor{urlcolor}{RGB}{255, 0, 0}
\hypersetup{
    colorlinks=true,
    linkcolor=linkcolor,
    citecolor=citecolor,
    urlcolor=urlcolor,
    linktoc=all,
    pdfborder={0 0 0}
}

\begin{document}

\title{Isospectrality and Operator Complexity}

\author{Pradip Kattel}
\email{pradip.kattel@unige.ch}
\affiliation{Department of Quantum Matter Physics, University of Geneva, Quai Ernest-Ansermet 24, 1211 Geneva, Switzerland}

\author{Yicheng Tang}
\affiliation{Department of Physics and Astronomy, Center for Materials Theory, Rutgers University,
Piscataway,  New Jersey, 08854, United States of America}

\author{Natan Andrei}
\affiliation{Department of Physics and Astronomy, Center for Materials Theory, Rutgers University,
Piscataway,  New Jersey, 08854, United States of America}

\begin{abstract}
We study a pair of exactly solvable, isospectral fermion chains, one strongly interacting and one quadratic, that nevertheless display remarkably different phase structures and operator dynamics. A nonlocal nonlinear unitary transformation maps one onto the other while preserving the entire many-body spectrum and converting local fermion operators into extended many-body strings. Thus, operators that evolve within a closed linear subspace in the quadratic model become interacting operators that generate increasingly higher-body terms and exhibit asymptotic Lanczos growth $b_n\propto\sqrt n$. Despite their identical spectra, the two models realize distinct phases and sharply different notions of operator complexity. Our results demonstrate that free many-body spectra and interacting operator dynamics are fundamentally compatible.
\end{abstract}

\maketitle

A central question in quantum many-body physics is how much information about a quantum system is encoded in its many-body spectrum. Spectral information underlies much of our understanding of equilibrium phenomena, including thermodynamics, response functions, and conventional classifications of phases of matter~\cite{sachdev1999quantum,altland2010condensed}. By contrast, developments in nonequilibrium quantum dynamics have shifted attention toward intrinsically dynamical notions such as quantum chaos~\cite{haake1991quantum}, operator spreading and scrambling~\cite{shenker2014black,maldacena2016bound,nahum2018operator,von2018operator}, out-of-time-order correlators~\cite{larkin1969quasiclassical,maldacena2016bound}, and Krylov complexity~\cite{parker2019universal,dymarsky2020quantum,rabinovici2021operator}. These developments suggest that quantum dynamics depends not only on spectral properties but also on the structure of operators under time evolution~\cite{nahum2018operator}. An open question is whether two local Hamiltonians with identical many-body spectra can nevertheless exhibit sharply different phase structures, operator growth, scrambling, and dynamical complexity.

In this work, we study two local fermionic Hamiltonians related by a nonlocal nonlinear fermion-to-fermion unitary transformation. Despite exact spectral equivalence, the two Hamiltonians realize distinct phases~\cite{tang2026topological} and qualitatively different operator dynamics, as revealed by the Liouvillian Lanczos coefficients and Krylov complexity~\cite{parker2019universal,dymarsky2020quantum,rabinovici2021operator}. Because the transformation is nonlocal, it reorganizes the local operator algebra, mapping local fermion operators into extended many-body strings. As a consequence, the two isospectral Hamiltonians exhibit distinct operator growth and scrambling dynamics~\cite{nahum2018operator}.

One of the Hamiltonians we consider is an interacting nearest-neighbor fermion chain, while the other is a quadratic Kitaev-type Hamiltonian. Unlike canonical interacting-free correspondences, which relate different microscopic degrees of freedom\footnote{
Canonical examples include the Jordan-Wigner solution of the Ising and XY chains~\cite{jordan1928paulische,lieb1961two}, the free-fermion point of the sine-Gordon model~\cite{luther1974backward,coleman1975quantum}, and more recent examples of ``free fermions in disguise''~\cite{fendley2019free}. Most such correspondences relate distinct microscopic degrees of freedom, such as spins, bosons, parafermions, and fermions.
}, the present construction maps an interacting fermion chain exactly onto a quadratic fermion model. Since the interacting model is unitarily equivalent to a quadratic free-fermion model, a broad class of observables admits exact Pfaffian representations determined by the propagator of the dual quadratic model~\cite{barouch1971statistical}. Local Fermionic operators evolve within a closed linear subspace in the quadratic theory, yielding bounded Lanczos coefficients, whereas repeated commutators in the interacting model generate increasingly higher-body operators, producing asymptotic Krylov growth $b_n\propto\sqrt n$. The same distinction appears in the out-of-time-ordered correlator (OTOC), where ballistic propagation coexists with nontrivial operator spreading.

\paragraph*{The Models:}
We consider an interacting nearest-neighbor chain of spinless fermions,
\begin{equation}
H_{\mathrm{int}}(\lambda)=H_K+\lambda H_{\mathrm{nn}},
\label{eq:Hint}
\end{equation}
consisting of the quadratic Kitaev chain,
$H_K=\sum_{j=1}^{N-1}\left[t\left(c_j^\dagger c_{j+1}+\mathrm{h.c.}\right)+\Delta_p\left(c_j c_{j+1}+\mathrm{h.c.}\right)\right]$
and a quartic density interaction term 
$H_{\mathrm{nn}}=\sum_{j=1}^{N-1}\left(2n_j-1\right)\left(2n_{j+1}-1\right),$     with $n_j=c_j^\dagger c_j $. Interacting extensions of the Kitaev chain have been studied previously using bosonization and numerical methods~\cite{stoudenmire2011interaction,gangadharaiah2011majorana}.
We restrict our attention to the dimerized point
$t=\Delta_p=1$,
so that $\lambda$ is the only tuning parameter in Eq.~(\ref{eq:Hint}).

Using the nonlocal and nonlinear fermion-to-fermion unitary transformation introduced in Ref.~\cite{tang2026topological},
\begin{align}
c_k^\dagger &=
\frac{i}{2}\left[\prod_{j=1}^{k-1}\left(1-2 f_j^\dagger f_j\right)^{k-1-j}\right]
\left[\prod_{l=1}^{k-1}\left(f_l+f_l^\dagger\right)\right]
\nonumber\\
&\quad\times
\left[\left(2 f_k^\dagger f_k-1\right)
-\left(\prod_{j=1}^{k-1}\left(1-2 f_j^\dagger f_j\right)\right)\left(f_k^\dagger-f_k\right)\right],
\label{eq:explicit_map}
\end{align}
the interacting Hamiltonian Eq.~\eqref{eq:Hint}  maps at the exactly dimerized point for arbitrary $\lambda$ to the quadratic nearest-neighbor Kitaev Hamiltonian, $H_{\mathrm{ff}}(\lambda)=U^{\dagger} H_{\mathrm{int }}(\lambda)U$, with\footnote{
The present construction provides a lattice analog of the Luther-Emery phenomenon, admitting both interacting and free-fermion representations of the same theory.
}

\begin{align}  
H_{\mathrm{ff}}(\lambda) = \sum_{j=1}^{N-1}
\Big[ (1+\lambda)\left( f_j^\dagger f_{j+1}
+ f_{j+1}^\dagger f_j \right) \nonumber\\
\qquad\qquad
+ (1-\lambda)\left( f_j^\dagger f_{j+1}^\dagger +
f_{j+1} f_j \right) \Big],
\label{eq:Hff_generic}
\end{align}
where the $f_j$ satisfy canonical anticommutation relations. The transformation maps a nearest-neighbor interacting fermion Hamiltonian exactly onto a nearest-neighbor quadratic Hamiltonian. At $\lambda=1$, the pairing term vanishes and Eq.~(\ref{eq:Hff_generic}) reduces to a nearest-neighbor tight-binding chain, even though the original $c$-fermion Hamiltonian remains interacting. At $\lambda=0$, the model becomes self-dual, with equal hopping and pairing amplitudes.
Equation~\eqref{eq:explicit_map} is generated by a nonlocal fermion-to-fermion unitary transformation $U$, satisfying $c_j = U f_j U^\dagger$ and $c_j^\dagger = Uf_j^\dagger U^\dagger$. Explicitly,
\begin{equation}
U=e^{-\frac{\pi}{4}\sum_{i=1}^{N}\left(\prod_{j<i}(1-2f_j^\dagger f_j)\right)
(f_i^\dagger-f_i)}e^{-i\frac{\pi}{2}\sum_{i=1}^{N}f_i^\dagger f_i}.
\label{eq:Ucf}
\end{equation}
The existence of $U$ guarantees exact isospectrality between $H_{\mathrm{int}}$ and $H_{\mathrm{ff}}$. Because $U$ contains Jordan-Wigner parity strings extending across the system, it is not a finite-depth local circuit and therefore maps local observables into extended many-body operators.
\begin{figure}
    \centering    \includegraphics[width=\linewidth]{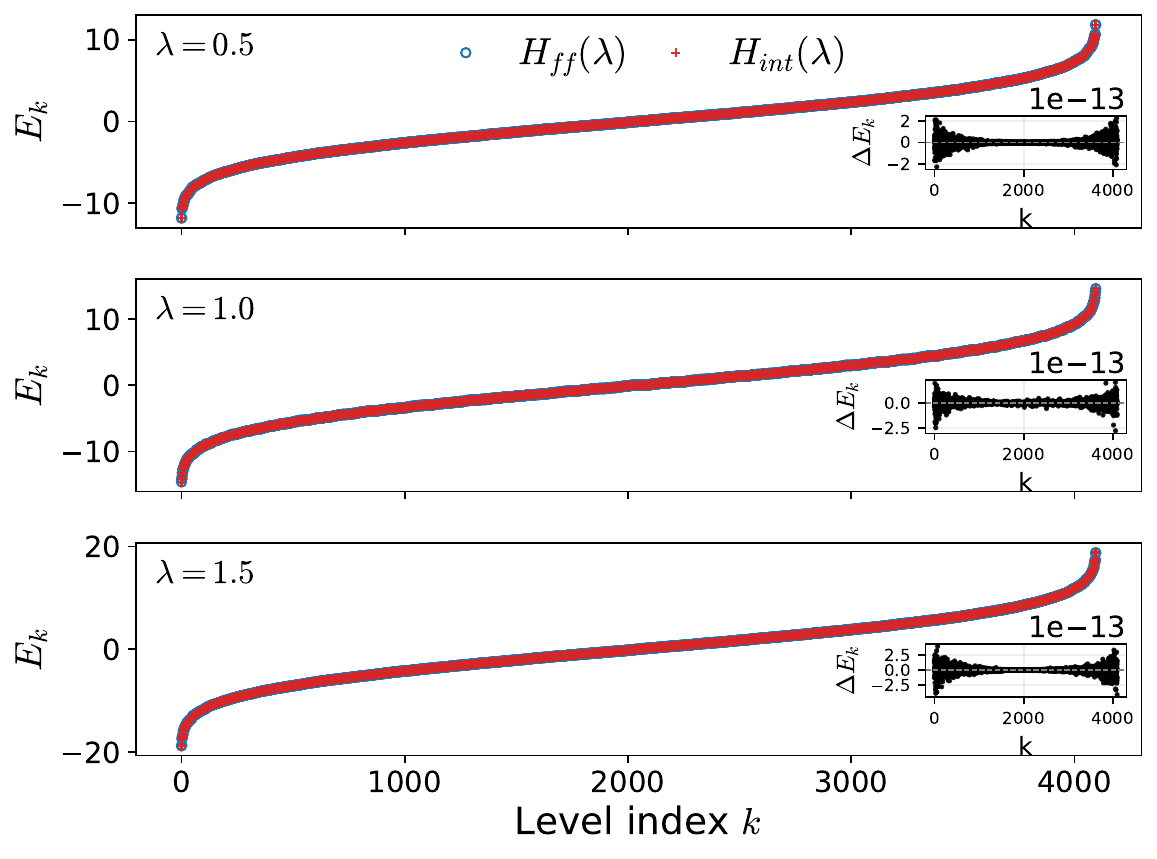}
   \caption{Many-body spectra of the quadratic Hamiltonian $H_{\mathrm{ff}}(\lambda)$ (blue circles) and the interacting Hamiltonian $H_{\mathrm{int}}(\lambda)$ (red crosses) for a chain of length $N=12$ at $\lambda=0.5,1,1.5$. The complete level-by-level agreement demonstrates the exact unitary equivalence between the interacting nearest-neighbor fermion chain Eq.~\eqref{eq:Hint} and the quadratic nearest-neighbor Hamiltonian Eq.~\eqref{eq:Hff_generic}. Inset: spectral differences $\Delta E_k=E_k^{(\mathrm{int})}-E_k^{(\mathrm{ff})}$ vanish to numerical precision.}
    \label{fig:placeholder}
\end{figure}

Having established isospectrality, we now examine the extent to which physical properties are constrained by the many-body spectrum. We begin with the phase structure of the two models and subsequently analyze dynamical diagnostics, including correlation functions, OTOCs, and Krylov complexity.

\paragraph*{The phase structure.}
Although the two Hamiltonians are isospectral and share the same bulk critical points at $\lambda=\pm1$, the phases on either side of these transitions are different. 
The quadratic Hamiltonian $H_{\mathrm{ff}}(\lambda)$ in Eq.~(\ref{eq:Hff_generic}) is a Kitaev chain in the BDI symmetry class, with hopping $t_{\mathrm{eff}}=1+\lambda$ and pairing $\Delta_{\mathrm{eff}}=1-\lambda$. Its quasiparticle gap closes at $\lambda=\pm1$, separating gapped phases characterized by winding numbers $\nu=+1$ for $|\lambda|<1$ and $\nu=-1$ for $|\lambda|>1$. Both phases support Majorana edge modes under open boundary conditions, while the change in winding number signals a topological phase transition at the gap-closing points~\cite{kitaev2009periodic,kitaev2001unpaired}.
However, the unitary transformation relating $H_{\mathrm{ff}}(\lambda)$ and $H_{\mathrm{int}}(\lambda)$ is highly nonlocal and is not a finite-depth local circuit~\cite{chen2010local}. The nature of the phases of the interacting Hamiltonian is markedly different, consisting of three phases separated by bulk gap closings at $\lambda=\pm1$~\cite{fidkowski2010effects,tang2026topological}: for $\lambda>1$ the system exhibits charge-density-wave order, for $-1<\lambda<1$ it lies in the SPT regime continuously connected to the Kitaev limit at $\lambda=0$, and for $\lambda<-1$ it enters a density-polarized regime. In the topological regime $-1<\lambda<1$, the interacting chain hosts two exponentially localized strong zero modes~\cite{fendley2016strong}, whose explicit forms are highly nonlocal many-body strings:
\begin{widetext}
\begin{align}
\Psi_L^{(c)} &= \mathcal N \sum_{m=0}^{\lceil N/2\rceil-1} (-\lambda)^m \left[
\left(\prod_{k=1}^{2m} \Big( \prod_{\ell=1}^{k-1}\big(1-2c_\ell^\dagger c_\ell\big)
\Big)\,(c_k+c_k^\dagger) \right)
\left(
\prod_{\ell=1}^{2m}\big(1-2c_\ell^\dagger c_\ell\big)
\right)\frac{c_{2m+1}-c_{2m+1}^\dagger}{i}
\right]\\
\Psi_R^{(c)}
&= \mathcal N \sum_{m=0}^{\lceil N/2\rceil-1} (-\lambda)^m \left[ \left(\prod_{k=1}^{N-2m-1}
\Big( \prod_{\ell=1}^{k-1}\big(1-2c_\ell^\dagger c_\ell\big) \Big)\,(c_k+c_k^\dagger) \right)
\Big(1-2c_{N-2m}^\dagger c_{N-2m}\Big)
\right],\quad \mathcal{N}=\sqrt{\frac{1-\lambda^2}{1-\lambda^{2\lceil N/2\rceil}}}.
\end{align}
\end{widetext}

\begin{figure*}
    \centering
    \includegraphics[width=\linewidth]{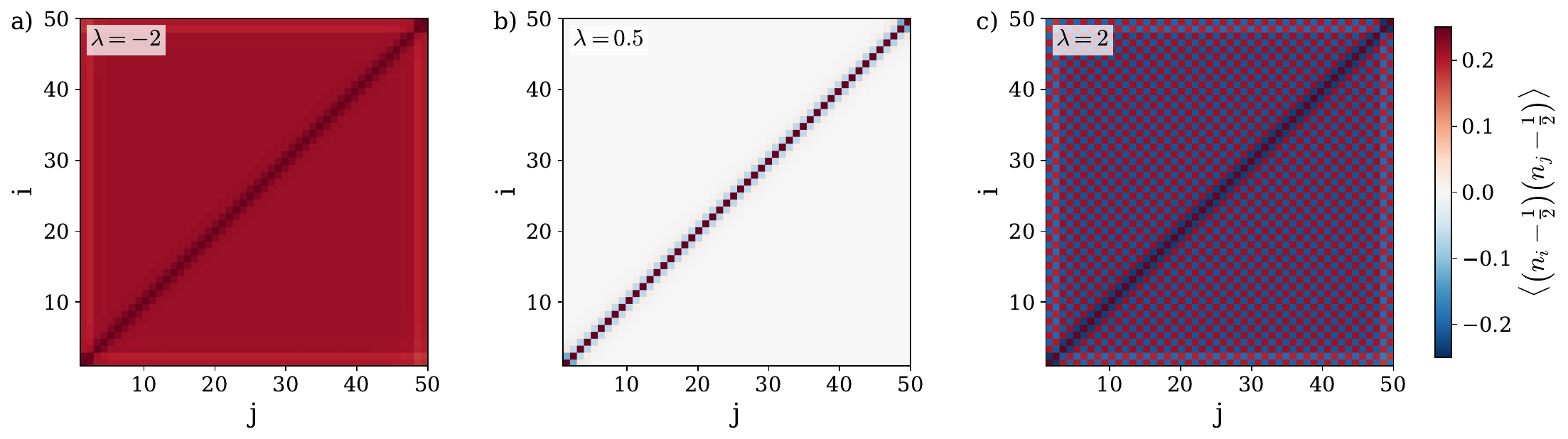}
    \caption{Ground-state density correlations 
$C_{ij}=\left\langle\left(n_i-\frac{1}{2}\right)\left(n_j-\frac{1}{2}\right)\right\rangle$ exhibit: 
(i) for $\lambda=-2$, a uniform long-distance value $C_{ij}\to m^2$ evaluated in the bulk (with $i,j$ far from the open boundaries), indicating density polarization without CDW order; 
(ii) for $\lambda=0.5$, exponential decay with separation, consistent with a gapped SPT phase lacking a local order parameter; and 
(iii) for $\lambda=2$, a checkerboard long-distance form $C_{ij}\to (-1)^{|i-j|} m^2$, diagnosing staggered CDW order, where $m^2=\sqrt{1-\frac{1}{\lambda^2}}$.}
    \label{fig:nncordd}
\end{figure*}

Unlike the simple Majorana edge operators of the quadratic model, the corresponding operators in the interacting chain are highly nonlocal many-body strings, illustrating how the duality reorganizes the local operator algebra and thereby modifies both phase structure and operator dynamics.

\paragraph*{Dynamics: Correlations and Operator Spreading}
The exact solvability of the models becomes especially transparent in a Majorana basis $\gamma_{2j-1}=f_j+f_j^\dagger,
\quad
\gamma_{2j}=i\left(f_j^\dagger-f_j\right)$, in terms of which the quadratic Hamiltonian $H_{\mathrm{ff}}(\lambda)$ becomes
\begin{equation}
H_{\mathrm{ff}}(\lambda)
=
i\sum_{j=1}^{N-1}
\left[
-\gamma_{2j}\gamma_{2j+1}
+
\lambda\,\gamma_{2j-1}\gamma_{2j+2}
\right].
\label{eq:H_majorana_local}
\end{equation}
Being quadratic in Majorana operators, the Heisenberg equations of motion of the model close linearly within the Majorana sector. Writing $\gamma=(\gamma_1,\dots,\gamma_{2N})^T$, the Hamiltonian may be expressed compactly as
\begin{equation}
H_{\mathrm{ff}}(\lambda)
=
\frac{i}{4}\gamma^TA(\lambda)\gamma,
\qquad
A(\lambda)^T=-A(\lambda),
\label{eq:H_majorana}
\end{equation}
where $A_{ab}(\lambda)$ denotes the coupling between Majorana modes
$\gamma_a$ and $\gamma_b$. The Heisenberg evolution then takes the form
\begin{equation}
\gamma_a(t)
=
\sum_{b=1}^{2N}
R_{ab}(t;\lambda)\gamma_b,
\qquad
R(t;\lambda)=e^{A(\lambda)t}.
\label{eq:single_particle}
\end{equation}
At infinite temperature,
$
\left\langle X \right\rangle_{\infty}
\equiv
2^{-N}\mathrm{Tr}(X),
$
Majorana contractions are determined exactly by the propagator $R(t;\lambda)$,
\begin{equation}
\left\langle \gamma_a(t_1)\gamma_b(t_2) \right\rangle_{\infty}
=
\left(R(t_1-t_2;\lambda)\right)_{ab}.
\label{eq:gamma_contr_infty}
\end{equation}
Introducing the antisymmetric contraction kernel
\begin{align}
\mathcal{G}_{ab}(t_1,t_2;\lambda)
&\equiv
\left\langle
\gamma_a(t_1)\gamma_b(t_2)
\right\rangle_{\infty}
-\delta_{ab},
\nonumber\\
\mathcal{G}_{ab}(t_1,t_2;\lambda)
&=
-\mathcal{G}_{ba}(t_2,t_1;\lambda),
\label{eq:Gkernel}
\end{align}
Wick's theorem reduces all even Majorana correlators to Pfaffians,
\begin{align}
\left\langle
\gamma_{i_1}(\tau_1)\cdots
\gamma_{i_{2m}}(\tau_{2m})
\right\rangle_{\infty}
=
\mathrm{Pf}(M),
\quad
M_{rs}
=
\mathcal{G}_{i_r i_s}(\tau_r,\tau_s;\lambda).
\label{eq:pf_general}
\end{align}

  We now turn to express the interacting  Hamiltonian $H_{\mathrm{int}}$ in terms of the Majorana variables.

Observing that the local parity operator can be written as $P_j\equiv 1-2f_j^\dagger f_j =-i\gamma_{2j-1}\gamma_{2j}$,
and introducing $S_k=\prod_{j=1}^{k-1}P_j^{k-1-j}$, the transformed fermion operator can be written as
\begin{equation}
c_k^\dagger
=
-\frac{1}{2}
\left(
\Gamma_{k,1}
+
\Gamma_{k,2}
\right),
\label{eq:ck-majorana-sum}
\end{equation}
where
\begin{align}
\Gamma_{k,1}
&=
iS_k
\left(\prod_{l=1}^{k-1}\gamma_{2l-1}\right)
P_k,
\label{eq:Gamma1}\\
\Gamma_{k,2}
&=
S_k
\left(\prod_{l=1}^{k-1}\gamma_{2l-1}\right)
\left(\prod_{j=1}^{k-1}P_j\right)\gamma_{2k},
\label{eq:Gamma2}
\end{align}
with each $\Gamma_{k,\alpha}$ being, up to an overall phase, a product of Majorana operators. Crucially, for fixed $k$, each local fermion operator is represented by only a finite number of Majorana monomials. This finite representability underlies the exact Pfaffian reduction of interacting correlators.
Although the interacting Hamiltonian is not quadratic in the physical $c$-fermion variables, its local operators are mapped by the nonlocal unitary transformation into finite sums of Majorana strings in the dual quadratic model. Let $\mathcal O$ be a local operator of the interacting model such that
\begin{equation}
U^\dagger \mathcal OU = \sum_{\alpha=1}^{M} \omega_\alpha\Gamma_\alpha, \qquad \Gamma_\alpha
= \gamma_{i_1(\alpha)}\cdots\gamma_{i_{\ell(\alpha)}(\alpha)},
\label{eq:O_expand}
\end{equation}
with finite $\ell(\alpha)$. Multi-time correlators of such operators reduce exactly to finite sums of Pfaffians determined entirely by the contraction kernel $\mathcal G_{ab}(t_1,t_2;\lambda)$ and hence by the propagator
$
R(t;\lambda)=e^{A(\lambda)t}
$
of the dual quadratic Hamiltonian. Consequently, equal-time correlators, dynamical correlators, and out-of-time-order correlators of local operators remain exactly computable, even though the physical $c$-fermion operators themselves are non-Gaussian composite Majorana strings and do not obey Wick factorization.

As a first application, we compute the ground-state density correlations of Eq.~\eqref{eq:Hint}, which provide a direct diagnostic of the three phases discussed above. In the bulk, translation invariance reduces the Pfaffian expressions to Toeplitz determinants~\cite{lieb1961two}, from which the long-distance behavior of
$ C_{ij} = \left\langle \left(n_i-\frac{1}{2}\right)
\left(n_j-\frac{1}{2}\right) \right\rangle $
follows as
$ C_{ij}\to m^2 $ in the density polarized phase ($\lambda<-1$),
$ C_{ij}\to0 $ in the topological superconductor phase ( $|\lambda|<1$),
and $ C_{ij}\to(-1)^{|i-j|}m^2 $
 in the CDW phase ($\lambda>1$ ), where $ m^2=\sqrt{1-\frac{1}{\lambda^2}}. $

\begin{figure}[h]
\centering
\includegraphics[width=\linewidth]{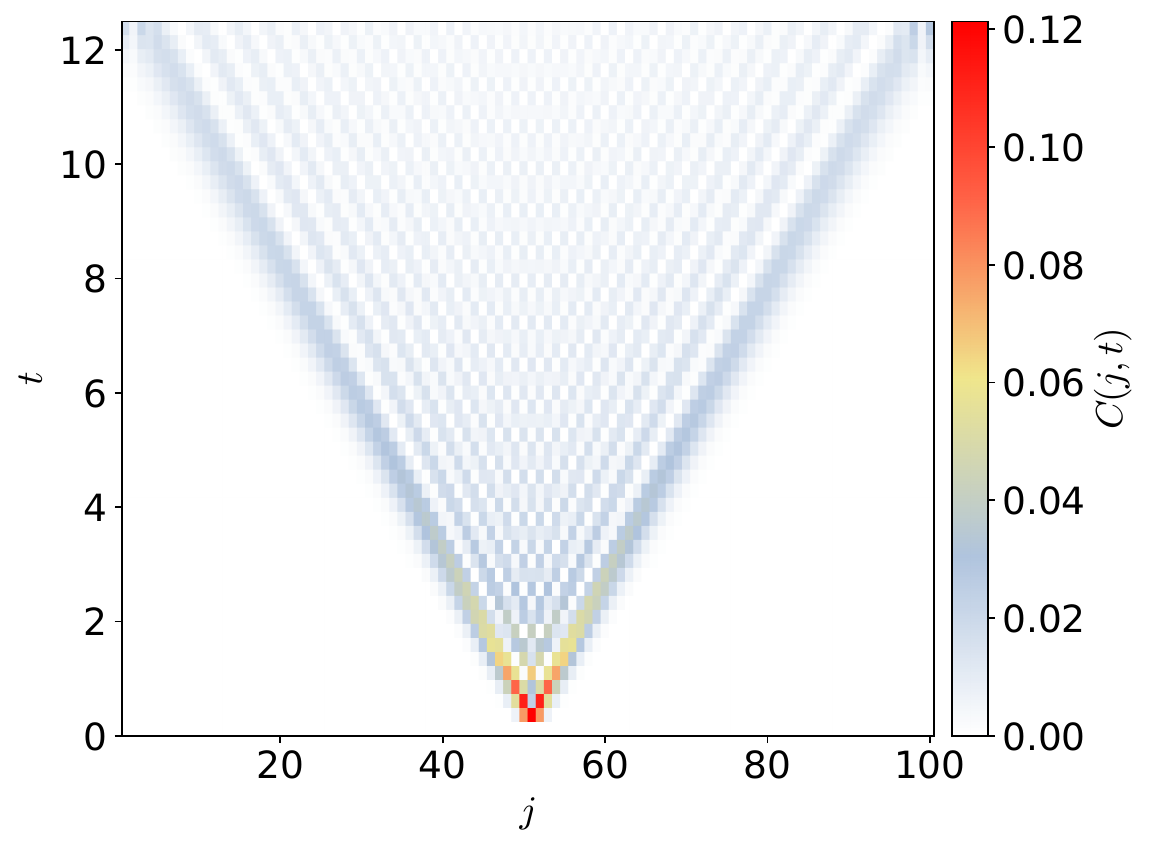}
\begin{tikzpicture}[remember picture,overlay]
\node at (-4.5,6.5) {\bfseries (a)};
\end{tikzpicture}
\includegraphics[width=\linewidth]{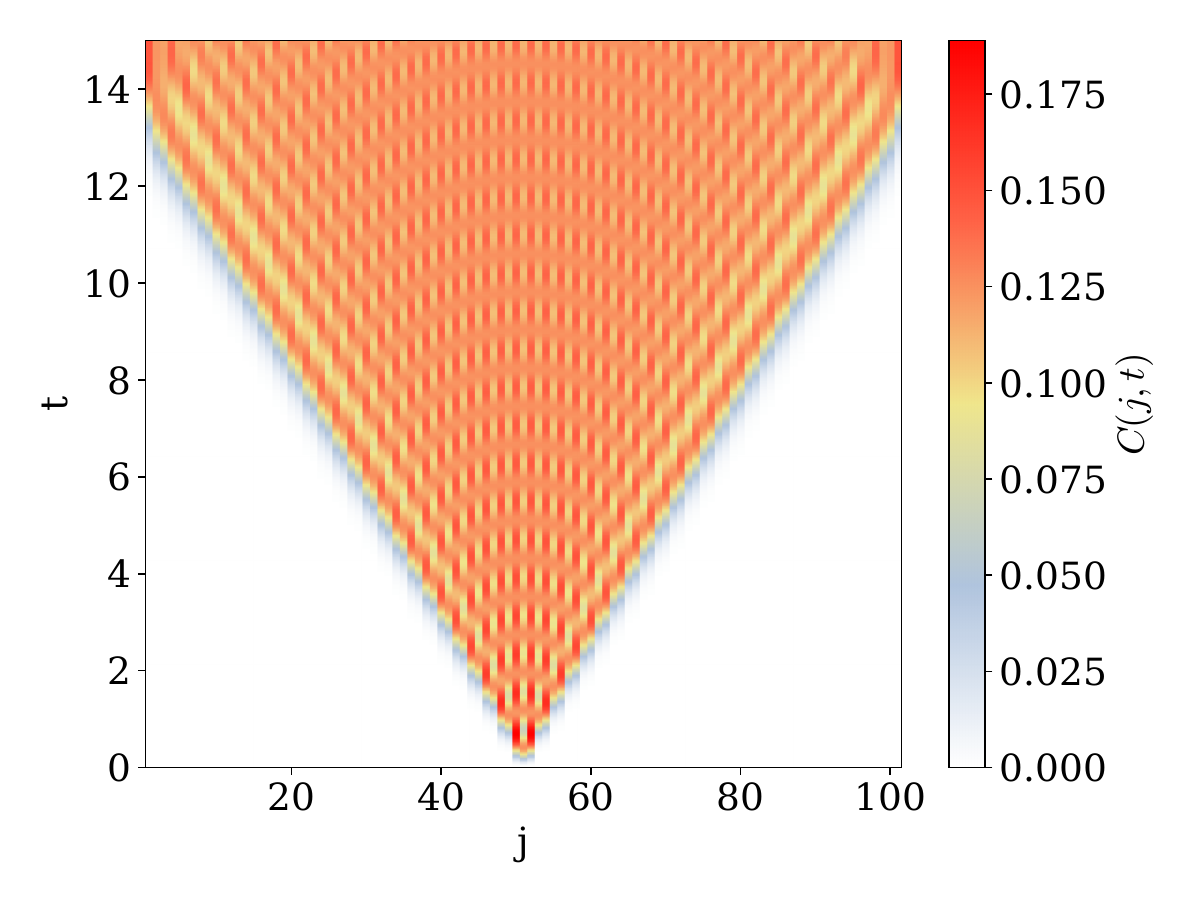}
\begin{tikzpicture}[remember picture,overlay]
\node at (-4.5,6.5) {\bfseries (b)};
\end{tikzpicture}
\caption{
Infinite-temperature OTOCs of centered density operators
$W_j=n_j-\frac{1}{2}$ and
$V=n_{j_0=50}-\frac{1}{2}$ at $\lambda=1$.
(a) OTOC of the quadratic dual Hamiltonian $H_{\mathrm{ff}}(1)$.
(b) OTOC of the interacting Hamiltonian $H_{\mathrm{int}}(1)$.
Both models exhibit the same ballistic light cone with
$v_{\max}=4$, reflecting their identical single-particle spectrum.
However, the interacting OTOC develops substantial weight throughout the light cone and a pronounced interference structure arising from the extended Majorana-string representation of local operators. The comparison illustrates that propagation velocities are determined by the spectrum, whereas operator growth depends on the operator algebra.
}
\label{fig:otoc}
\end{figure}
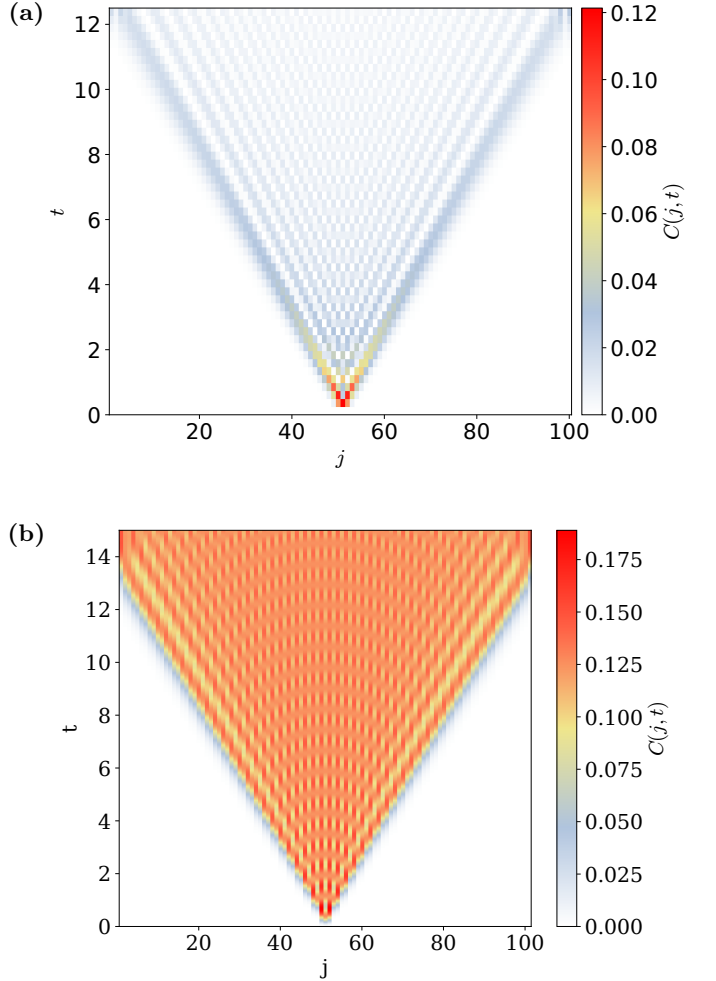

We next ask whether a system isospectral to a free-fermion model can nevertheless exhibit nontrivial operator spreading. To address this question, we consider the infinite-temperature OTOC built from centered densities $W_j \equiv n_j-\frac{1}{2}$ and $V \equiv n_{50}-\frac{1}{2}$ with
$W_j(t) \equiv e^{iH_{\mathrm{int}}t}W_j e^{-iH_{\mathrm{int}}t}$
defined as~\cite{maldacena2016bound}
\begin{align}
C(j,t) &\equiv \frac{1}{2}\left\langle \left[W_j(t),V\right]^\dagger \left[W_j(t),V\right]
\right\rangle_{\infty}.
\label{eq:C_vs_F}
\end{align}
The numerical results in Fig.~\ref{fig:otoc}(a,b) are shown for $\lambda=1$, where the dual Hamiltonian $H_{\mathrm{ff}}(1)$ is a tight-binding chain with dispersion $\varepsilon(k)=4\cos k$ and maximal velocity $v_{\max}=4$. The OTOC exhibits a ballistic front propagating at exactly this velocity, demonstrating that the light cone is determined by the common free-fermion spectrum. Yet the interior of the light cone develops a rich interference pattern absent for free density operators [Fig.~\ref{fig:otoc}(a)], reflecting the fact that physical densities are mapped to extended Majorana strings by the nonlocal unitary transformation [Fig.~\ref{fig:otoc}(b)]. In this case, the ballistic front velocity is inherited from the common free-fermion quasiparticle spectrum, whereas the interior structure reflects operator growth.

\paragraph*{Isospectrality versus operator growth: Krylov complexity.}
The OTOC probes spatial spreading, but does not directly characterize growth within the operator Hilbert space. To probe this aspect of dynamics, we turn to Krylov complexity. Because the nonlocal unitary transformation reorganizes the operator algebra, the two isospectral Hamiltonians act very differently on operator space, leading to sharply distinct Lanczos dynamics. We quantify this through the Lanczos recursion of the Liouvillian $\mathcal{L}(\mathcal{O})\equiv[H,\mathcal{O}]$, using the
infinite-temperature inner product $\left(\mathcal{A},\mathcal{B}\right)=2^{-N}\mathrm{Tr}\left(\mathcal{A}^\dagger\mathcal{B}\right)$.
Starting from a normalized seed $\mathcal{O}_0$ and setting $\mathcal{O}_{-1}=0$, define the \emph{unnormalized}
residual and its norm
\begin{align}
\widetilde{\mathcal{O}}_{n+1} &\equiv \mathcal{L}\mathcal{O}_n-b_n\mathcal{O}_{n-1},
\label{eq:lanczos_step}\\
b_{n+1} &\equiv \sqrt{\left(\widetilde{\mathcal{O}}_{n+1},\widetilde{\mathcal{O}}_{n+1}\right)},
\quad \mathcal{O}_{n+1}\equiv \frac{1}{b_{n+1}}\widetilde{\mathcal{O}}_{n+1}.
\label{eq:bn_def}
\end{align}
The Lanczos recursion also allows a diagonal coefficient
$a_n=\left(\mathcal{O}_n,\mathcal{L}\mathcal{O}_n\right)$.
For the commutator Liouvillian and the infinite-temperature
Hilbert-Schmidt inner product, $a_n=0$ in the Hermitian operator
sector, yielding Eqs.~(\ref{eq:lanczos_step})-(\ref{eq:bn_def}).

\begin{figure}[b]
\centering
\includegraphics[width=\columnwidth]{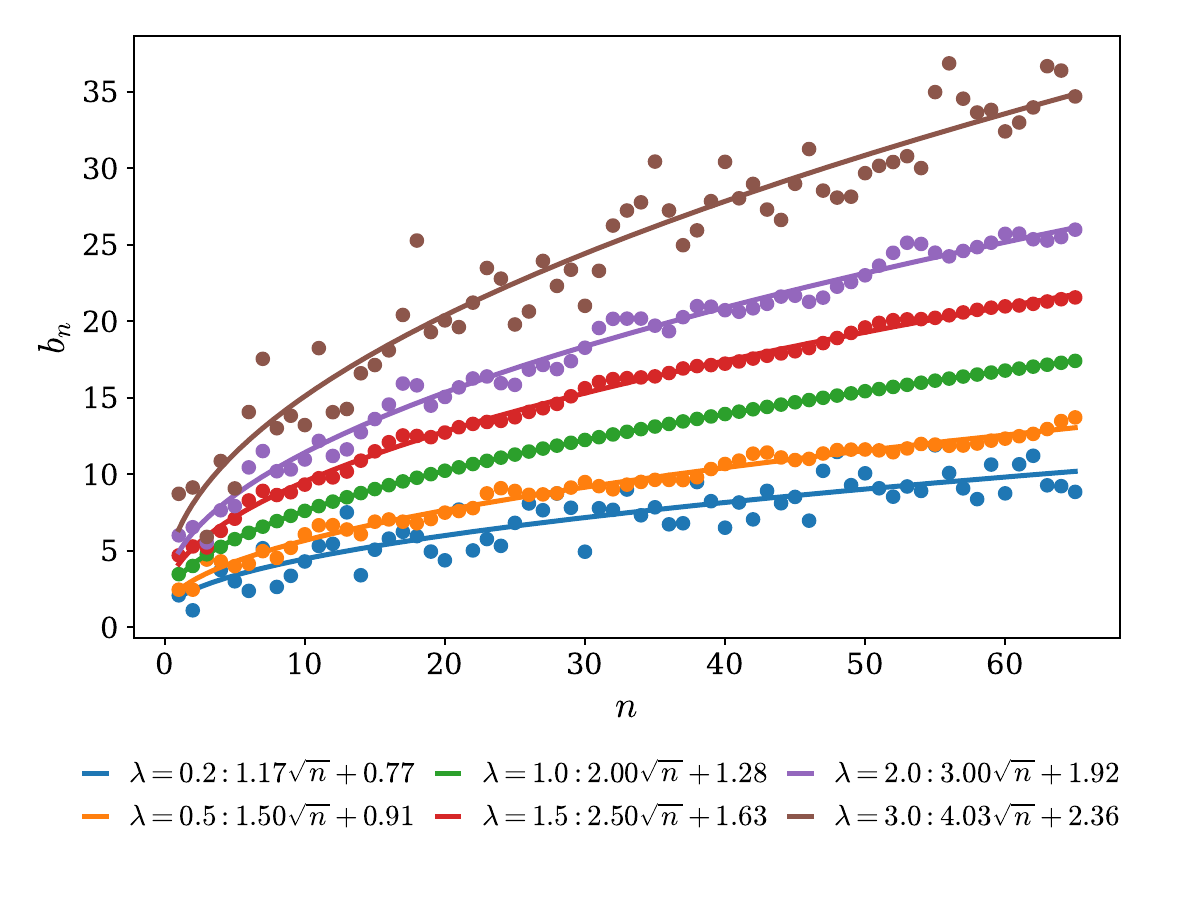}
\caption{Lanczos coefficients $b_n$ from the Liouvillian Krylov recursion with a local Majorana seed, computed numerically for a chain of length $N=100$ up to $n=65$ for various values of $\lambda$. To suppress short-range numerical fluctuations, the raw sequence is smoothed using a Savitzky–Golay filter (window size 11, quadratic polynomial)~\cite{savitzky1964smoothing}. The resulting coefficients are consistent with
$b_n\propto \sqrt{n}$ over the accessible Krylov depths,
demonstrating distinct operator complexity despite exact isospectrality.}
\label{fig:bn}
\end{figure}

For the quadratic Hamiltonian $H_{\mathrm{ff}}(\lambda)$, choosing $\mathcal{O}_0$ to be a local Majorana, the commutator closes within the linear Majorana sector, equivalently $\mathcal{L}\gamma_a=\sum_b K_{ab}(\lambda)\gamma_b$ with real antisymmetric $K(\lambda)$. In the thermodynamic limit with a bulk seed and for Krylov depths $n$ well below the distance to the boundaries, the Lanczos coefficients approach a period-two form,
\begin{equation}
b_n^{\mathrm{ff}} \simeq 
\begin{cases}
2\sqrt{1+\lambda^2}, & n=1,\\[4pt]
2, & n\ge 2\ \text{and}\ n\ \text{odd},\\[4pt]
2|\lambda|, & n\ge 2\ \text{and}\ n\ \text{even},
\end{cases}
\qquad (H=H_{\mathrm{ff}}(\lambda)),
\label{eq:bn_alt}
\end{equation}
with nonuniversal $O(1)$ deviations at small $n$ that depend on the choice of seed. For a finite chain of length $N$, the Krylov subspace generated from a single-Majorana seed is contained in the $2N$-dimensional linear Majorana sector, so the Lanczos recursion necessarily terminates, i.e.\ there exists $n_\ast\le 2N$ such that $b_{n_\ast+1}=0$.

For the interacting Hamiltonian $H_{\mathrm{int}}$, the Lanczos amplitudes grow sublinearly with Krylov depth; in the bulk thermodynamic limit one expects the asymptotic form $b_n \sim \alpha(\lambda)\sqrt{n}$ as $n\to\infty$~\cite{parker2019universal}, where $\alpha(\lambda)$ is an even, model-dependent prefactor. Heuristically, $b_{n+1}^2$ measures the component of $\mathcal L O_n$ that explores previously inaccessible directions in operator space. In a local interacting chain, the active support of $O_n$ grows roughly linearly with Krylov depth, so the number of independent commutator processes scales as $\mathcal N_n\propto n$. Assuming approximate orthogonality of these contributions gives $b_{n+1}^2\propto n$, and hence $b_n\propto\sqrt n$.

The first few Lanczos amplitudes can be computed exactly and are found to depend only on $\lambda^2$, implying $b_n(\lambda)=b_n(-\lambda)$. Explicit expressions for these coefficients are provided in the End Matter. To investigate the behavior at larger Krylov depths, we computed Lanczos sequences numerically for the interacting model at various values of $\lambda$. The resulting data exhibit a pronounced intermediate regime displaying approximately $\sqrt n$
growth of the Lanczos coefficients (Fig.~\ref{fig:bn}), suggesting convergence toward the asymptotic form
$b_n\sim\alpha(\lambda)\sqrt n$.

\textit{Conclusion:} We have used an exactly solvable pair of isospectral Hamiltonians to identify which aspects of many-body physics are not determined by the spectrum alone. Although the two models share the same many-body eigenvalues, they realize distinct phase structures, different local operator algebras, and sharply different operator-growth dynamics. The underlying mechanism is a nonlocal fermion-to-fermion unitary transformation that preserves the spectrum while strongly reorganizing the local operator algebra. As a consequence, the interacting Hamiltonian exhibits Lanczos growth consistent with $b_n\sim\alpha(\lambda)\sqrt n$, whereas the dual quadratic Hamiltonian has bounded Lanczos coefficients. These results show that spectral equivalence alone does not determine phase structure, local operator algebra, or operator complexity. More broadly, they demonstrate that the many-body spectrum alone is insufficient to characterize the dynamical content of a quantum system, and that free many-body spectra and operator dynamics exhibiting characteristic interacting signatures can coexist.

\acknowledgments

\textit{Acknowledgments:} We thank Colin Rylands and Adrian B. Culver for helpful discussions and insightful comments. This work was supported by the Swiss National Science Foundation under Division II (Grant No.~200020-219400).
\bibliography{ref}

\section*{End Matter}

The first few Lanczos amplitudes can be computed exactly from the recursion relations in Eqs.~(\ref{eq:lanczos_step})-(\ref{eq:bn_def}). Defining
$B_n(x)\equiv b_n^2(\lambda),
\qquad x=\lambda^2$, the first five coefficients are
\begin{widetext}
\begin{align}
B_1(x) &= 4(2x+1), \quad  B_2(x) = \dfrac{16x(x+2)}{2x+1}, \quad B_3(x) = 4\,\dfrac{20x^2+21x+10}{2x^2+5x+2}, \nonumber\\
B_4(x) &= 2\,\dfrac{128x^4+362x^3+873x^2+618x+128}{20x^3+61x^2+52x+20}, \nonumber\\
B_5(x) &= 2\,\dfrac{2048x^6+40656x^5+151274x^4+209413x^3+131686x^2+53256x+4608}
{1280x^5+4324x^4+11009x^3+11652x^2+6308x+1280},
\label{eq:bn_squared_exact}
\end{align}
\end{widetext}
where $b_n(\lambda)=\sqrt{B_n(\lambda^2)}\qquad(n=1,\dots,5)$. These expressions depend only on $\lambda^2$ and satisfy $b_n(\lambda)=b_n(-\lambda)$. 
\end{document}